# VIBRO-ACOUSTIC MODULATION BASELINE-FREE NON-DESTRUCTIVE TESTING

Dimitri Donskoy[*] and Dong Liu

*Stevens Institute of Technology, Hoboken, NJ 07030 United States*

[*]*Corresponding author:* <ddonskoy@stevens.edu>

Vibro-Acoustic Modulation method for detection and characterization of various structural and material flaws has been actively researched for the last two decades. Most of the studies focused on detection and monitoring of macro-cracks requiring well established baseline (no-damage) value of the modulation index. The baseline value is specific for a particular structure, measuring setup, and other factors and can't be established in many practical situations without a long term monitoring looking for a relative change in the Modulation Index. In this work, we propose and investigate a baseline-free Vibro-Acoustic Modulation method, which does not require monitoring of relative Modulation Index change, unlike conventional approach. It was hypothesized that the nonlinear mechanisms (and respective nonlinear response) of a structure are different for undamaged and damaged material. For example: material without damage or at early stages of fatigue have classic elastic or hysteretic/dissipative nonlinearity while damaged (cracked) material may exhibit contact bi-linear or Hertzian nonlinear mechanisms. These mechanisms yield different power law dependencies of Modulation Index ($MI$) as function of applied vibration frequency input amplitude, $B$: $MI \sim B^\beta$. Thus, quadratic nonlinearity yields linear dependence, $\beta = 1$, and Hertzian nonlinearity results in $\beta < 1$. Other nonlinear mechanisms yield different power laws. Therefore, measuring power damage coefficient $\beta$ instead of $MI$ may offer testing without established reference value. It also offer some insights into the nonlinear mechanisms transformation during damage evolution. This approach was experimentally investigated and validated.

1.  **Introduction**

   Vibro-Acoustic Modulation (VAM) technique has been introduced in 1990s [1], [2] for detection of contact-type defects such as cracks and delaminations. Later, the method was applied to monitoring a damage evolution at the microscopic level demonstrating its high sensitivity to damage initiation before macro defects are developed, [3]. There have been numerous follow up studies, for example [4], [5], [6], [7], [8] of the method applied to a variety of structural and material defects demonstrating high damage sensitivity of VAM as well as its other advantageous features. A comprehensive review of the VAM related publications is given in [9].

   VAM utilizes nonlinear interaction (modulation) of a high frequency ultrasonic wave (carrier signal) having frequency $\omega$ and a low frequency vibration (modulating vibration) with frequency $\Omega \ll \omega$. Material nonlinearity and especially highly nonlinear damage-related interfaces such as cracks, disbonds, as well as structural contact interfaces (bolted connections, overlays, etc.) cause the wave interaction/modulation. In majority of studies, the modulation is quantified by a Modulation Index (MI) defined in the spectral domain as the ratio of the side-band spectral components at frequencies $\omega \pm \Omega$ to the amplitude of the carrier. MI does not differentiate between the type and origin of the modulating cause: be it material, damage, or structural-related nonlinearities. It is assumed that the material and structural nonlinearities do not change over life of the structure, thus setting up a **baseline** MI value for undamaged structure. Damage, developed at some point, increases MI over its baseline, indicating damage presence and severity: the higher MI, the greater the damage. Therefore, in its basic form, VAM needs an established baseline value. This works well for **monitoring** of damage evolution (monitoring a relative change in MI), for example in Structural Health Monitoring (SHM) applications. For the non-destructive testing applications, however, the baseline value is not always available and, often, may not be determined.

   There are a few publications refereeing to "baseline-free VAM" [10], [11], [12] . Examination of these papers reveals that the authors assumed that VAM is inherently baseline-free method because in the absence of the damage – there is no modulation, therefore the baseline is zero. In practice, however, it is far from zero due to material and measurements setup nonlinearities as well as structural (non-damage) nonlinearities such as structural contact interfaces.

   In this study, we propose physics-based baseline-free VAM testing exploiting the differences in nonlinear mechanisms at difference stages of damage evolution.

## 2. Nonlinear mechanisms and their VAM manifestations

Development of nonlinear acoustic non-destructive testing such as harmonic, frequency mixing, and modulation methods, [13], stimulated active studies of related physical nonlinear mechanisms. A comprehensive review of classical and non-classical nonlinear acoustic models is given by Broda, et. al.,[14]. Besides classical nonlinear elasticity, there is a variety of so-called non-classical nonlinear mechanisms: contact acoustic nonlinearities (CAN), hysteresis, thermo-elasticity, and nonlinear dissipation. All of these mechanisms contribute to acoustic nonlinear interactions. Here we focus on one particular interaction between high frequency ultrasonic waves and low frequency vibrations, which is utilized in VAM method. Specifically, we are interested in mostly overlooked effect of MI dependence on the amplitudes of the interacting signals for different nonlinear mechanisms.

For two-wave interaction, these dependences are different: classical quadratic nonlinearity of stress-strain Hooke's law yields linear dependence of combination frequencies amplitude on interacting signal amplitudes, [2]:

$$A_\pm \sim A*B \ , \qquad (1)$$

where $A$ and $B$ are amplitudes of the high frequency ($\omega$) ultrasound and the low frequency ($\Omega$) vibration, respectively, $A_\pm$ are amplitudes of the spectral components at the combination frequencies $\omega \pm \Omega$. The modulation index $MI$, defined as

$$MI = \frac{A_- + A_+}{2A} \ , \qquad (2)$$

is independent of the high frequency amplitude, $MI(A)$ = constant, and linear proportional to the amplitude of the low frequency vibration: $MI \sim B$.

Non-classical nonlinear mechanisms may manifest themselves with different amplitude dependences. Knowing these dependencies may help to identify the respective nonlinear mechanism and to develop a baseline-free testing methodology.

Some of the models yield theoretically predicted dependences, such as the above-mentioned quadratic model, while others, mathematically more complicated, do not easily reveal such dependencies. Here we will use numerical simulations using model's strain-stress relationship in generic scalar formulation, $\sigma(\varepsilon)$, for high and low frequency harmonic strain inputs:

$$\varepsilon = A\cos(\omega t) + B\cos(\Omega t) ,\tag{3}$$

and computing spectral amplitudes at the combination frequencies. This approach is not a full modelling of wave interactions, as it does not take into account many effects such as wave propagation and resonances, kinematic nonlinearity, mode conversions, vector (tensor) nature of interacting fields, etc. It provides, however, a simple way to predict the amplitude dependencies at the *source* of the nonlinear interaction defined by nonlinear constitutive equation, $\sigma(\varepsilon)$, even for very complex models. The above-mentioned unaccounted phenomena may mask or distort these source dependencies, so additional efforts will be needed (and discussed later) to recover/unmask the source amplitude dependencies.

Below, we consider a few examples of source nonlinear mechanisms and the resulting *MI* amplitude dependencies. The results of the modelling will be presented as a power function of Load Ratio $B_i/B_j$:

$$MI_i/MI_j \sim (B_i/B_j)^\beta ,\tag{4}$$

where $MI_i$ and $MI_j$ are the *MI*s defined by the equation (2) for the input LF amplitudes $B_i$ and $B_j$. where $i \neq j$. For example, for amplitudes $B_1, B_2, B_3, B_4, B_5$: $B_i/B_j = B_2/B_1, B_3/B_1, B_3/B_2, B_4/B_2$, etc.

### 2.1 Classical quadratic nonlinear elasticity

Well-studied quadratic nonlinearity is described by a quadratic term in Taylor's expansion of the Hooke's law:

$$\sigma = L\varepsilon - N\varepsilon^2 \tag{5}$$

where $L$ and $N$ are the linear and nonlinear elastic coefficients, respectfully.
Substitution of Eq.(3) into Eq.(5) reveals the amplitude dependence of Eq.(1). This well-known result can be used to verify our MatLab code to be used for more complex models. Indeed, as expected, MatLab computed dependence of normalized *MI* vs. *A* and *B* amplitudes (also normalized), shown in Fig.1, demonstrate theoretically predicted amplitude dependencies with power coefficient $\beta = 1$.

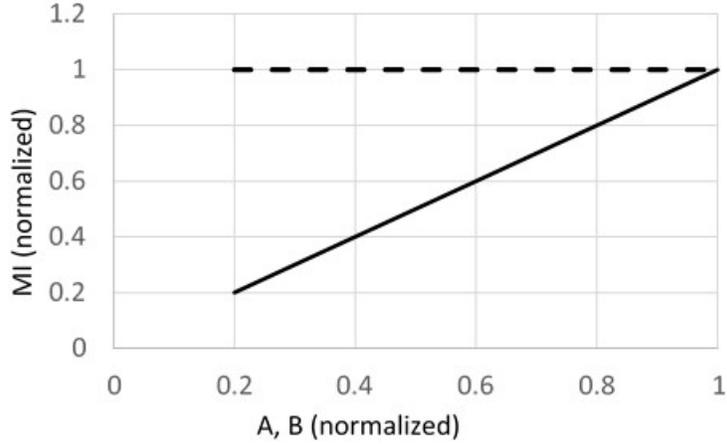

Fig.1. *MI* vs. *B* (solid) and *MI* vs. *A* (dashed).

## 2.2 Bi-linear contact acoustic nonlinearity

Bi-linear stress-strain dependence, Fig.2a, Eq.(6), at the contact interfaces was introduced in 1980s to model cracks in beams[15], [16] and disbonds/delaminations, [17], for a relatively high strains leading to the opening and closing of the interface.

It is worthwhile to notice that the bi-linear model yields only even harmonics ($2\Omega$, $4\Omega$, $6\Omega$, …) so the modulation spectrum contains the side-band components at frequencies $\omega \pm \Omega$, $\omega \pm 2\Omega$, $\omega \pm 4\Omega$, … as shown in Fig.2c. The bi-linear model power coefficient, determined by the Eq.(4), $\beta = 0$, that is: the normalized Modulation Index does not depend on the relative increase in LF amplitude B. With this, the modulation index, *MI* ~ *N/L*.

$$\sigma = L\varepsilon - N|\varepsilon| = \begin{cases}(L - N)\varepsilon, & \varepsilon \geq 0 \\ (L + N)\varepsilon, & \varepsilon < 0\end{cases}, \tag{6}$$

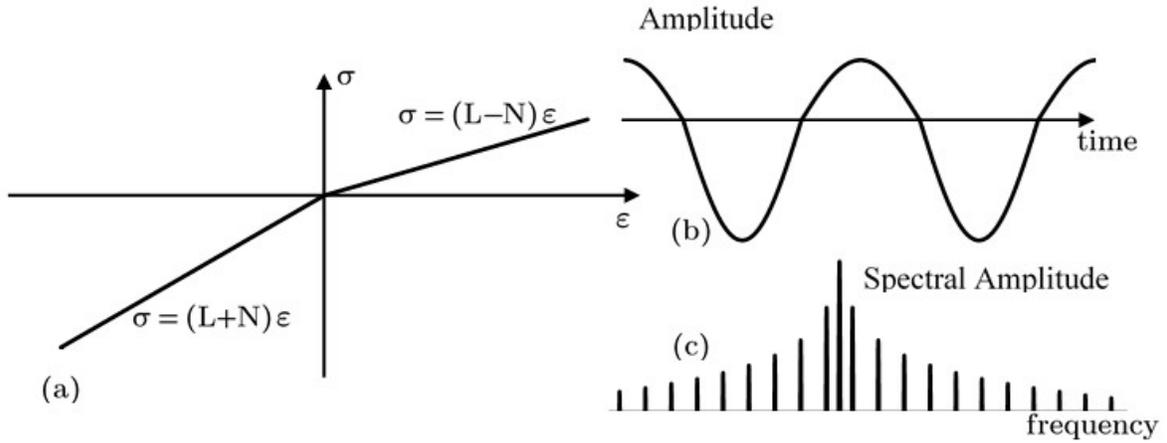

Fig.2. Case of bi-linear stiffness: (a) – stress-strain, (b) – LF output waveform, (c) – HF modulated spectrum.

### 2.3 Rough-surface contact acoustic nonlinearity

More realistic model of the contact interfaces, especially for a lower strain, is a rough-surface contact in which the curved asperities on both sides of the interface are in contacts and deformed under the dynamic stresses. The deformation could be elastic, plastic, or their combination. The deformation changes the contact area with the complex strass-strain relationships, which are dependent on the shape and size of the contacting asperities and other conditions such as slip, friction, adhesiveness, etc. For non-adhesive frictionless elastic deformation, the following stress-strain equation can be used:

$$\sigma = L\varepsilon - N\varepsilon^S, \qquad (7)$$

In this model, we modify the input signal (3) as following

$$\varepsilon = \varepsilon_0 + A\cos(\omega t) + B\cos(\Omega t), \qquad (8)$$

where $\varepsilon_0$ is constant strain: $\varepsilon_0 > A+B$. Under this condition, the total strain is always positive; therefore, there is no separation of the contacts.

The power coefficient $\beta$ of Eq.(4) depends on the power $S$ and the ratio of nonlinear/linear coefficients $N/L$. Assuming the spherical shape of the asperities, the power $S = 1.5$. In this case, Fig.3 illustrates stress-strain and normalized $MI(B)$ dependences for various $N/L$ ratios.

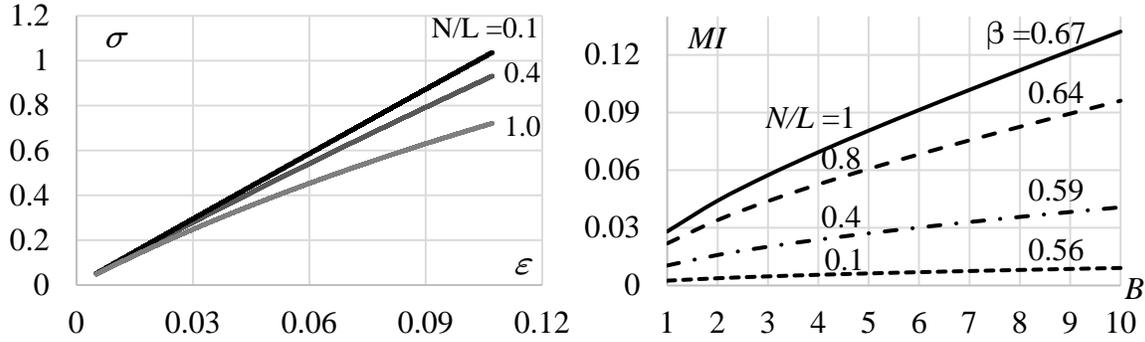

Fig.3. Left: Stress-strain dependence, Right: $MI$ vs $B$ for various $N/L$ ratios. All units are arbitrary normalized.

As this example demonstrates, the power coefficient $\beta$ varies within the range $0.5 – 0.7$ depending on the ratio of linear and nonlinear coefficients. In real life, $\beta$ variability could be even larger, due to variability of the coefficients $S$, $N$, $L$ and the combined effect of other nonlinear mechanisms. For example, for $S = 2.5$ and $N/L = 0.1$, the power coefficient $\beta = 1.65$.

### 2.4 Hysteretic nonlinearity

Nonlinear hysteretic behaviour in various solid material has been observed experimentally in numerous studies, for example [18], [19], [20], [21], [22]. Observations of acoustic nonlinear manifestations (amplitude-dependent attenuation, resonance frequency shift on acoustic amplitude, and others) in micro-inhomogeneous solids such as rocks, "soft" metals (zinc, copper), fatigues materials, etc. are explained using hysteretic nonlinearity [21], [22], [23]. Although physical mechanisms of the hysteretic behaviour are still debated, there are many phenomenological models has been proposed. To illustrate the effect of the hysteresis nonlinearity on MI(B) dependence we use the model first proposed by Nazarov, et.al.,[21]:

$$\sigma = \begin{cases} L\varepsilon - N_1\varepsilon^2\,; & \varepsilon > 0,\ \dot{\varepsilon} > 0 \\ L\varepsilon - N_1 B\varepsilon\,; & \varepsilon > 0,\ \dot{\varepsilon} < 0 \\ L\varepsilon + N_2\varepsilon^2\,; & \varepsilon < 0,\ \dot{\varepsilon} < 0 \\ L\varepsilon - N_2 B\varepsilon\,; & \varepsilon < 0,\ \dot{\varepsilon} > 0 \end{cases} \qquad (9)$$

Here the input strain is given by Eq.(3) assuming $B >> A$. It is interesting to consider two scenarios: symmetrical ($N_1 = N_2$) and asymmetrical ($N_1 \neq N_2$) hysteresis, Fig. 4.

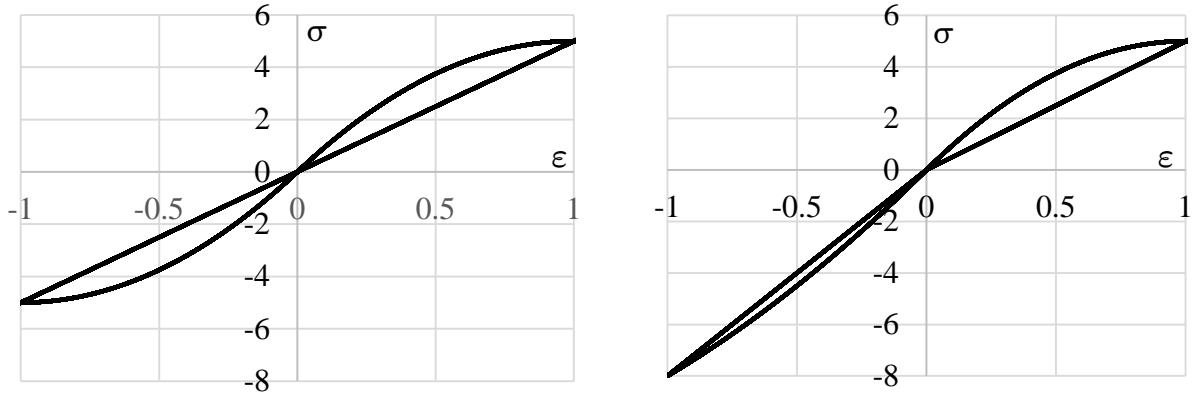

Fig.4 Symmetrical ($N_1/N_2 =1$, left) and asymmetrical ($N_1/N_2 =2.5$, right) hysteresis. All units are arbitrary normalized.

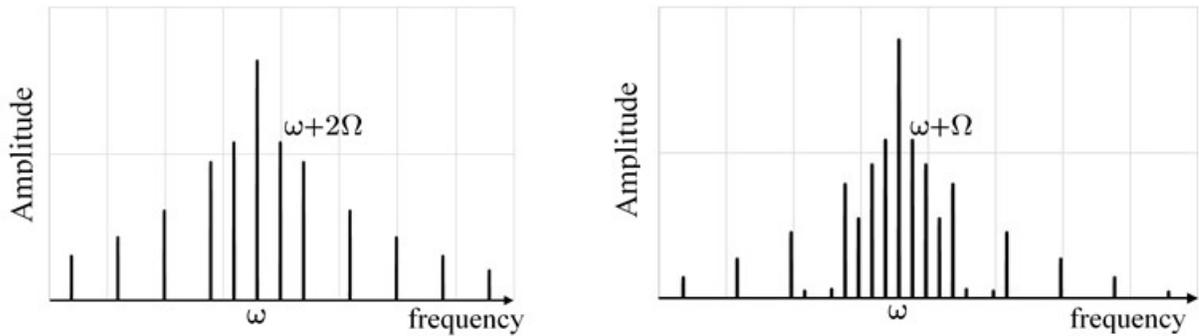

Fig.5. Modulation spectra for symmetrical (left) and for asymmetrical (right) hysteretic dependences of Fig. 4.

Modulation spectra for the above hysteretic dependencies are shown in the Fig.5 demonstrating that only non-symmetrical hysteresis yield modulation spectral components at the frequencies $\omega \pm \Omega$ which we are looking for. It was emphasized, [21], that asymmetrical hysteresis is more realistic as it reflects an asymmetrical nature of the compression vs. tension processes.

The power coefficient, $\beta$, for asymmetrical hysteresis depends on the combination of linear and nonlinear parameters, $L$, $N_1$, and $N_2$ and varies from 1.0 to 1.5.

It should be noted that there are wide variations of non-classical nonlinear models and their combinations, which would not be possible (and is not necessary) to discuss within the frame of the present work. The above examples confirm the hypothesis that the different nonlinear mechanisms yield different power coefficients β that may vary in range from 0 to 1.7 or more. The next question to answer is: if the power coefficient measurements could be served as a reliable and robust indication of damage condition and its evolution. Only experimental testing can answer this question.

## 3. Experimental investigation of *β* during fatigue damage evolution

The experimental verification of the proposed hypothesis was conducted on a number of A108 steel bars measuring 25.4 mm x 2.54 mm x 3.175 mm (10" x 1" x 1/8") subjected to tensile fatigue 10 Hz, 20 kN cycling using 810 MTS machine. In the centre of the bar, there is 0.635 mm (¼") diameter hole, so the stress and respective damages were concentrated between the hole and the edges of the bar. The fatigue cycles run until the breakage of the bar, as shown in Fig.6, which typically happened after ~ 100 thousand cycles.

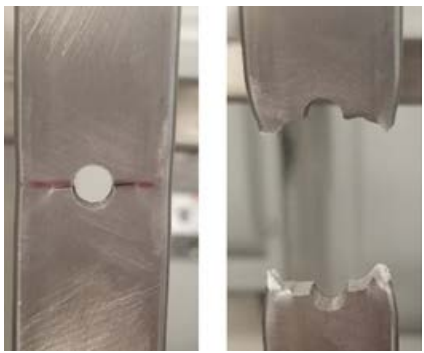
Fig.6. Test bar stress area.

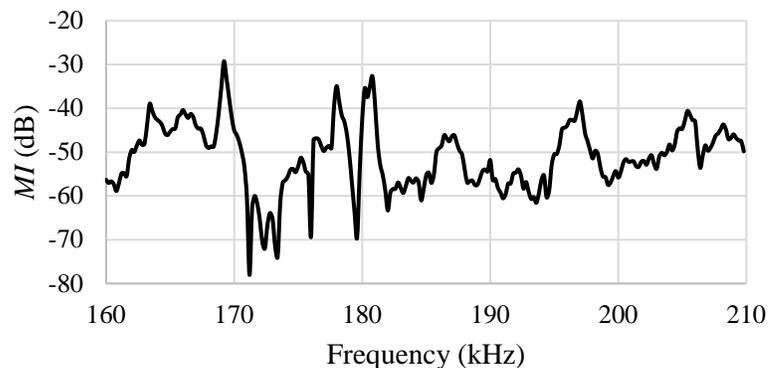
Fig.7. *MI* vs Frequency.

During the fatigue cycling, approximately after every 5000 cycles 20kN fatigue cycling were switched to a lower range of 0.5, 1, 1.5, 2, and 2.5 kN in succession. This 10 Hz low range cycling was used as a modulating vibration with respective amplitudes $B_i$ = 0.5 to 2.5 kN. Simultaneously, the high frequency ultrasonic signal was injected into the bar and received with a pair of piezo-ceramic transducers epoxy-glued 3 inches apart with the hole in the middle. The ultrasonic signal was step-swept across a wide frequency range of 120 kHz to 200 kHz with 0.5 kHz step. At each frequency the *MI* was measured and recorded. The example of the recorded *MI* vs. frequency is shown in Fig.7 demonstrating high variability of *MI* with the frequency. This variability, reported in many publications, is due to wave propagation, reflections/resonances, mode conversion, etc. within the bar and is difficult to account for, especially in real structures with complex geometry.

Instead, *MI* averaging across the wide frequency range provides reliable estimate of the structure nonlinearity, has been well documented, [3], [13]. Fig.8 shows averaged *MI*s across the frequency range vs normalized fatigue life of one of the tested samples. *MI*s are measured for five LF amplitudes: $B_i$ = 0.5, 1.0, 1.5, 2.0, and 2.5 kN showing onset of the fatigue damage at app. 80% - 90 % of the sample fatigue life.

The top solid line in Fig.8 is the power coefficient *β* (with scale on the right axis) calculated from these *MI*s using power trendline (regression) fitting as shown in Fig.9. The *β* curve clearly correlates with *MI* damage curves showing the damage onset at ~ 80-90%. This proves that the power coefficient is indeed follows the change in mechanisms of nonlinearity: here for the background nonlinearity (between 20% and 70% of the fatigue life, β is within the range 1.5 – 1.7.

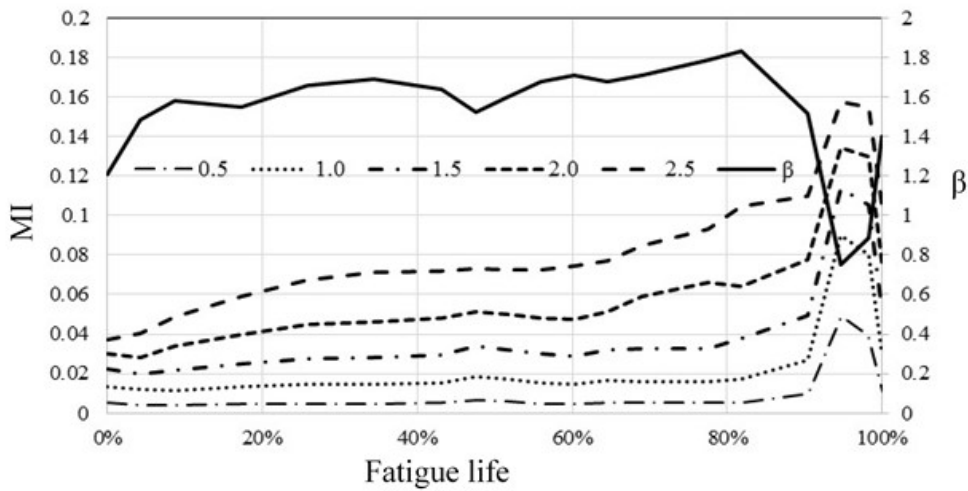

Fig.8. Averaged *MI* for five LF modulating amplitudes ($B_i$ = 0.5, 1.0,…2.5kN) vs % of fatigue life. Top solid line is calculated power coefficient *β* based on these five *MI* dependences.

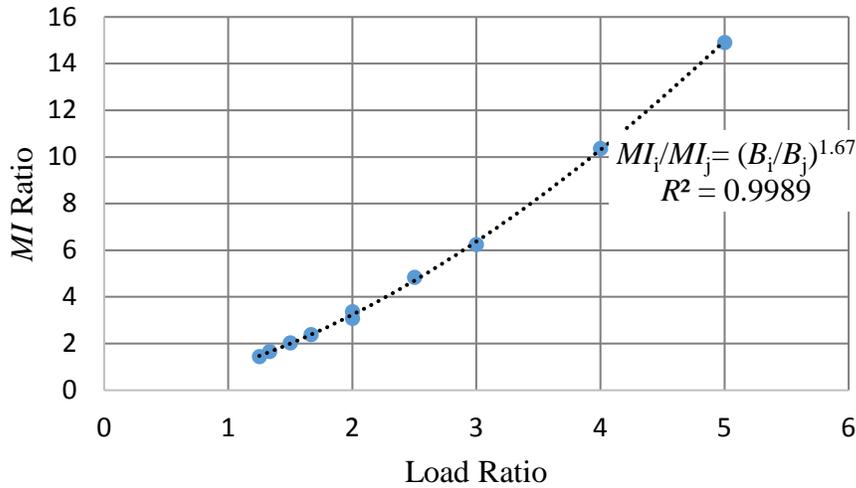

Fig.9. Example of *MI* Ratio vs Load Ratio with fitted power trendline (dashed) showing $\beta$ = 1.67 determined with high reliability (the coefficient of determination $R^2$ = 0.9989).

This pattern repeats itself for multiple samples as shown in Fig.10. It demonstrates very tight range of 1.5 – 1.6 before the onset of damage with significant drop with the development of macro-cracks (above 90% of the life). Here $\beta$ variation during the initial 10% of the life is likely due to setup settling (tightening the grip connections, etc).

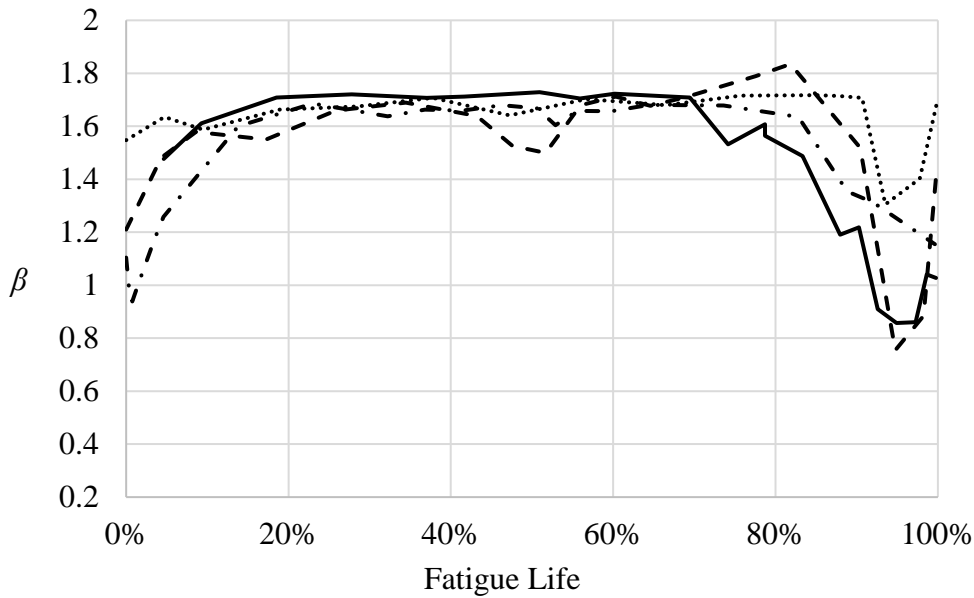

Fig.10. Power Coefficient $\beta$ vs. Fatigue life for four identical A108 steel samples.

## 4. Static Load Correction

The experimental results, demonstrated in the previous chapter, support our hypothesis that *MI* vs. vibration amplitude dependence expressed as a power law can be used as in indicator of changing nonlinear mechanisms, thus damage evolution indicator. The absolute value of the power damage coefficient $\beta$ should be associated with a particular nonlinear mechanism. One would expect that in the undamaged samples the main source of nonlinearity is a week material nonlinearity described by quadratic term in the constitutive equation (5). This should render $\beta = 1$ for the undamaged sample, while our test shows $\beta \sim 1.6 - 1.7$. This discrepancy brought our attention to effect of static component of the load used in the test. Fig.11 shows the waveforms of the applied vibrations with amplitudes $B_i$ = 0.5, 1.0, 1.5, 2.0, and 2.5 kN. It also shows that each applied waveform contains a corresponded static component force: $F_i$ = 0.75, 1.0, 1.25, 1.5, and 1.75kN which were necessitated by the operation of the tensile stress machine.

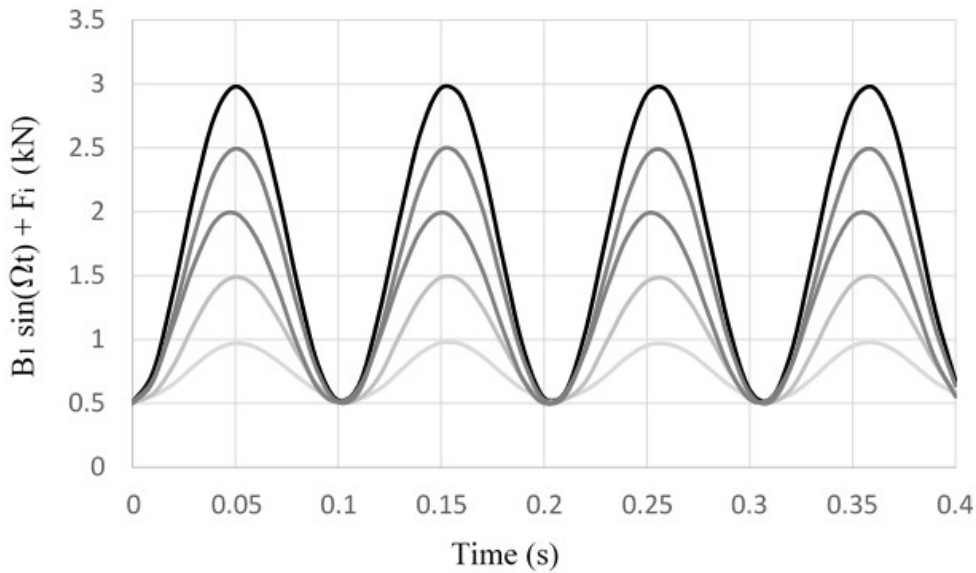

Fig.11. Waveforms of the applied 10 Hz vibrations with amplitudes $B_i$ and corresponded static force $F_i$ .

It is well known, [24], that the static stress increases the manifestation of the nonlinear acoustic signals. This effect has been utilized to determine nonlinear parameters of solids as well as to measure residual static stresses and is known as acousto-elasticity, [25]. In our experiments, the static stress is different for each vibration level and, therefore, its effect on the nonlinear measurements must be accounted and corrected for. Respectively, we modified the Eq.(4) as follows:

$$MI_i/MI_j \sim (B_i F_i/B_j F_j)^{\beta c}, \tag{9}$$

where $\beta c$ is the corrected (for effect of static load $F_i$) power damage coefficient.

Fig. 12 shows corrected $\beta c$ derived from the data of Fig.10. Remarkably, as anticipated, $\beta c$ is very close to 1 during the undamaged portion of the fatigue life for all tested samples.

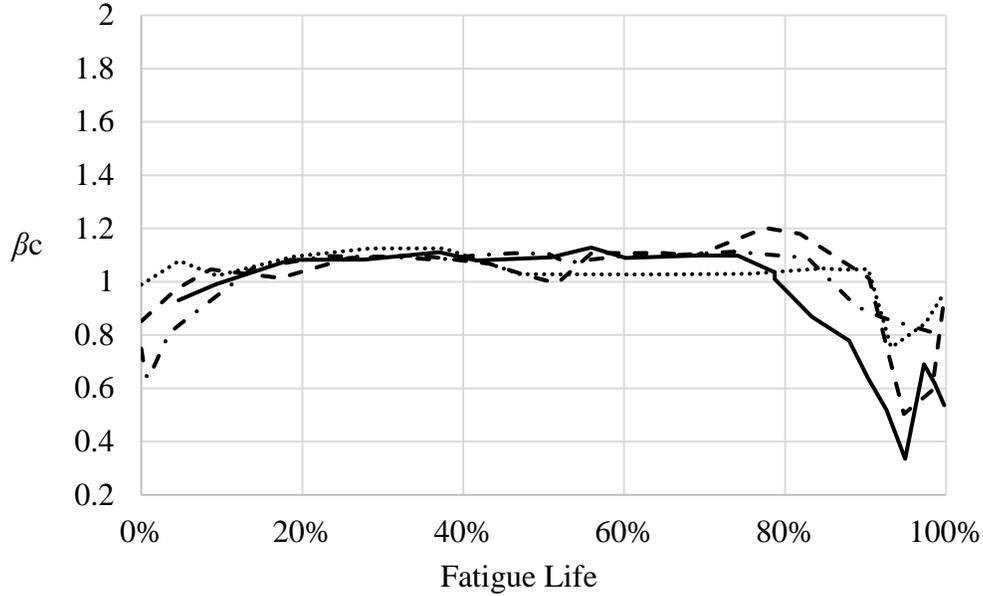

Fig.12. Corrected for the static stress power damage coefficient $\beta c$ vs fatigue life for four A108 steel bars.

5.  **Comparison with Conventional Ultrasonic and Eddy Current Baseline-free testing**

Fig. 12 demonstrate the ability of the proposed VAM baseline-free approach to detect early state damage at approximately 80% of the fatigue life. It is interesting to compare this with the conventional baseline-free techniques, such as ultrasonic (UT) and eddy-current (EC). To do this, we used off-the-shelf EPOCH 650 Digital Ultrasonic Flaw Detector equipment from Olympus America, Inc. For eddy current (EC), a NORTEC 600 Eddy Current Flaw Detector equipment (also from Olympus America, Inc.) was employed. To achieve the highest sensitivity to incipient fatigue damage, the highest available frequency probes were utilized for these tests: 20 MHz, 0.125 inch diameter ultrasonic delay line probe and 12 MHz, 0.125 inch diameter eddy current probe. The UT and EC measurements were conducted during the fatigue test over the area of the anticipated damage accumulation along the line shown in Fig.6. Both systems were able to detect the initial damage at 85% - 88% of the fatigue life, and a crack became visible above 90% of the fatigue life. This is yet another confirmation of high sensitivity of VAM technique to incipient damage detection as was reported in [3].

Comprehensive comparison of VAM and conventional acoustic emission (AE) method has been recently reported in [26]. Similarly, it showed VAM incipient damage detection at 80% versus AE damage detection at 85% of the fatigue life.

## 6. DISCUSSION AND CONCLUSION

The proposed baseline-free VAM non-destructive testing approach is based on the hypothesis that nonlinear mechanisms responsible for the vibro-acoustic modulation are different before and after the damage. The modelling and simulation revealed that there are noticeably different power damage coefficients $\beta$, Eq.(4), for various nonlinear mechanisms (NM). The expectation was that prior to the damage, the material exhibits classic nonlinear behaviour described by the quadratic nonlinearity, Eq.(5). As the modelling shown, this would yield the power coefficient $\beta = 1$. Indeed, the experimental results, Fig.12, show the power coefficient very close to unity. As damage accumulates and is developed into a macro-crack, the nonlinear mechanism should change yielding different power coefficients as summarized in the Table 1. Note, that there are other NMs proposed and discussed in the literature, see review paper [9], which are not modelled here.

Table 1. Nonlinear mechanisms and corresponding power coefficients

| Nonlinear Mechanism | Power Coefficient $\beta$ |
| --- | --- |
| Classical Quadratic, Eq.(5) | 1 |
| Bi-Linear, Eq.(6) | 0 |
| Contact (Herzian) Nonlinearity, Eq.(7) | 0.5 – 0.7 |
| Asymmetric Hysteretic, Eq.(9) | 1 – 1.5 |

This theoretical prediction is corroborated by the experiment, Fig.12, showing $\beta$ falling below unity as damage became apparent through observation of the increase in *MI*. It appeared that for the fatigued-damaged samples, the change (reduction) of $\beta$ is conducive with the contact nonlinear mechanism rather than bi-linear or hysteretic nonlinearities.

Unlike Modulation Index, which is a relative measure of nonlinearity irrespective of the NM, power damage coefficient $\beta$ is an absolute measure specific to a particular NM. Therefore, $\beta$-measurements, corrected for a static component of applied low frequency stress, $\beta c$, may offer baseline-free damage detection as opposite to *MI*-measurements suitable mostly for damage evolution monitoring. Initial experimental results, summarized in Fig.12 and corroborated with the theoretical predictions, are encouraging and support the proposed baseline-free damage detection approach. In these tests $\beta c \sim 1$ indicating undamaged material quadratic nonlinearity before the developed fatigue damage, and $\beta$ drops to below 1 as damage is evolved into a macro-crack. These below unity $\beta c$ values are conducive with a contact-type nonlinearity associated with crack interface.

Comparison with the conventional baseline-free methods: ultrasonic, eddy current, and acoustic emission tests, demonstrate higher sensitivity of baseline-free VAM to incipient fatigue damage detection.

## ACKNOWLEDGMENT

This work was supported by New Jersey Department of Transportation [contract #17-60125].